\documentclass [] {aa} 
\usepackage{epsfig, graphicx,longtable}
\usepackage{natbib}
\usepackage{times}

\bibpunct{(}{)}{;}{a}{}{,}

\begin{document}


\title{Beryllium anomalies in solar-type field stars\thanks{Based 
    on observations collected with the VLT/UT2 Kueyen telescope 
    (Paranal Observatory, ESO, Chile) using the UVES spectrograph 
    (Observing runs 66.C-0116\,A, 66.D-0284\,A, and 68.C-0058\,A), and 
    with the William Herschel and Nordic Optical Telescopes, operated at 
    the island of La Palma by the Isaac Newton Group and jointly by Denmark,
    Finland, Iceland, and Norway, respectively, in the Spanish Observatorio 
    del Roque de los Muchachos of the Instituto de Astrof\'{\i}sica 
    de Canarias.}}


\author{N.C.~Santos\inst{1,2} \and 
        G.~Israelian\inst{3} \and 
	S.~Randich\inst{4} \and
        R.J.~Garc\'{\i}a L\'opez\inst{3,5} \and
	R.~Rebolo\inst{3,6}
	}

\offprints{Nuno C. Santos, \email{Nuno.Santos@oal.ul.pt}}

\institute{
        Centro de Astronomia e Astrof{\'\i}sica da Universidade de Lisboa,
        Observat\'orio Astron\'omico de Lisboa, Tapada da Ajuda, 1349-018
        Lisboa, Portugal
     \and
	Observatoire de Gen\`eve, 51 ch.  des 
	Maillettes, CH--1290 Sauverny, Switzerland
     \and
	Instituto de Astrof{\'\i}sica de Canarias, E-38200 
        La Laguna, Tenerife, Spain
     \and
        INAF/Osservatorio Astrofisico di Arcetri, Largo Fermi 5, I-50125
        Firenze, Italy
     \and	
        Departamento de Astrof\'{\i}sica, Universidad de La Laguna,
        Av. Astrof\'{\i}sico Francisco S\'anchez s/n, E-38206, La
        Laguna, Tenerife, Spain
     \and
        Consejo Superior de Investigaciones Cient\'{\i}ficas, Spain 
        }
	
\date{Received / Accepted } 

\titlerunning{Be anomalies in solar-type field stars} 


\abstract{
We present a study of beryllium (Be) abundances in a large sample of
field solar-type dwarfs and sub-giants spanning a large range of effective 
temperatures. The Be abundances, computed using a very
uniform set of stellar parameters and near-UV spectra obtained
with 3 different instruments, 
are used to study the depletion of this light element. 
The analysis shows that Be is severely depleted
for F stars, as expected by the light-element depletion models.
However, we also show that Beryllium abundances decrease
with decreasing temperature for stars cooler than $\sim$6000\,K, a result
that cannot be explained by current theoretical models including rotational mixing,
but that is, at least in part, expected from the models that take into 
account internal wave physics. In particular, the light element 
abundances of the coolest and youngest stars in our sample suggest that Be, 
as well as lithium (Li), has already been burned early during their evolution. Furthermore, we find strong evidence for the existence of a Be-gap for solar-temperature 
stars. The analysis of Li and Be abundances in the sub-giants of our sample also shows the
presence of one case that has still detectable amounts of
Li, while Be is severely depleted. Finally, we compare the derived Be abundances with Li abundances
derived using the same set of stellar parameters. This gives us the 
possibility to explore the temperatures for which the onset of 
Li and Be depletion occurs.
}

\maketitle

\section{Introduction}

The light elements lithium (Li), beryllium (Be), and boron (B)
are important tracers of the internal stellar structure and kinematics. 
Since they are destroyed at relatively low temperatures, they give us
an idea about how the material inside stars is mixed with the
hotter interior. Their analysis, together or separately, can thus give us important 
information about the mixing and depletion processes. 

\begin{table*}
\caption[]{Derived Be abundances (BD and HD numbers up to 80\,000).}
\begin{tabular}{lcccrrcccccr}
\hline
Star     & T$_{\mathrm{eff}}$ & $\log{g}_{spec}$ & $\xi_{\mathrm{t}}$ & \multicolumn{1}{c}{[Fe/H]} & $\log{N(Be)}$ & $\sigma(Be)$ & Instr.$^\dagger$ & S/N & $v\,\sin{i}$ & source$^{\dagger\dagger}$ & $\log{N(Li)}$\\
         & [K]                &  [cm\,s$^{-2}$]  &  [km\,s$^{-1}$]    &                            &               &              &        &     & [km\,s$^{-1}$]&       &              \\
\hline
Sun$^{\star}$      &5777    &4.44    &1.00    &0.00    &1.10    &0.05    &--      &--      &1.9     &(e)      &1.10 \\
\object{BD\,-10\,3166}	&5320	 &4.38    &0.85    &0.33    &$<$ 0.50  &--    &[2]     &20	&1.58	 &(a)	  &--     \\
\object{HD\,   870}	&5447	 &4.57    &1.13    &-0.07   & 0.80   & 0.15   &[1]     &130	&1.77	 &(a)	  &$<$ 0.20 \\
\object{HD\,  1461}	&5768	 &4.37    &1.27    &0.17    & 1.14   & 0.13   &[1]     &120	&1.71	 &(a)	  &$<$ 0.51 \\
\object{HD\,  1581}	&5956	 &4.39    &1.07    &-0.14   & 1.15   & 0.11   &[1]     &140	&2.16	 &(a)	  & 2.37  \\
\object{HD\,  3823}	&5948	 &4.06    &1.17    &-0.25   & 1.02   & 0.12   &[1]     &130	&1.99	 &(a)	  & 2.41  \\
\object{HD\,  4391}	&5878	 &4.74    &1.13    &-0.03   & 0.64   & 0.11   &[3]     &150	&2.72	 &(a)	  &$<$ 1.09 \\
\object{HD\,  6434}	&5835	 &4.60    &1.53    &-0.52   & 1.08   & 0.10   &[3]     &150	&1.30	 &(a)	  &$<$ 0.85 \\
\object{HD\,  7570}	&6140	 &4.39    &1.50    &0.18    & 1.17   & 0.10   &[3]     &180	&3.82	 &(a)	  & 2.91  \\
\object{HD\,  9826}	&6212	 &4.26    &1.69    &0.13    & 1.05   & 0.13   &[6]     &120	&9	 &(b)	  & 2.55  \\
\object{HD\, 10647}	&6143	 &4.48    &1.40    &-0.03   & 1.19   & 0.10   &[3]     &150	&4.87	 &(a)	  & 2.80  \\
\object{HD\, 10697}	&5641	 &4.05    &1.13    &0.14    & 1.31   & 0.13   &[5]     &40	&--	 &--	  & 1.96  \\
\object{HD\, 10700}	&5344	 &4.57    &0.91    &-0.52   & 0.83   & 0.11   &[3]     &180	&0.90	 &(a)	  &$<$ 0.41 \\
\object{HD\, 12661}	&5702	 &4.33    &1.05    &0.36    & 1.13   & 0.13   &[5]     &40	&--	 &--	  &$<$ 0.98 \\
\object{HD\, 13445}	&5163	 &4.52    &0.72    &-0.24   &$<$ 0.40  &--      &[1]     &150	&1.27	 &(a)	  &$<$-0.12 \\
\object{HD\, 14412}	&5368	 &4.55    &0.88    &-0.47   & 0.80   & 0.11   &[3]     &190	&1.42	 &(a)	  &$<$ 0.44 \\
\object{HD\, 16141}	&5801	 &4.22    &1.34    &0.15    & 1.17   & 0.13   &[1]     &120	&1.95	 &(a)	  & 1.11  \\
\object{HD\, 17051}	&6252	 &4.61    &1.18    &0.26    & 1.03   & 0.13   &[1]     &150	&5.38	 &(a)	  & 2.66  \\
\object{HD\, 19994}	&6190	 &4.19    &1.54    &0.24    & 0.93   & 0.12   &[3]     &140	&8.10	 &(a)	  & 1.99  \\
\object{HD\, 20010}	&6275	 &4.40    &2.41    &-0.19   & 1.01   & 0.10   &[3]     &180	&4.63	 &(a)	  & 2.13  \\
\object{HD\, 20766}	&5733	 &4.55    &1.09    &-0.21   &$<$-0.09  &--      &[3]     &200	&1.98	 &(a)	  &$<$ 0.97 \\
\object{HD\, 20794}	&5444	 &4.47    &0.98    &-0.38   & 0.91   & 0.11   &[3]     &250	&0.52	 &(a)	  &$<$ 0.52 \\
\object{HD\, 20807}	&5843	 &4.47    &1.17    &-0.23   & 0.36   & 0.11   &[3]     &160	&1.74	 &(a)	  &$<$ 1.07 \\
\object{HD\, 22049}	&5073	 &4.43    &1.05    &-0.13   & 0.80   & 0.13   &[5]     &100	&2.13	 &(a)	  &$<$ 0.25 \\
\object{HD\, 22049}	&5073	 &4.43    &1.05    &-0.13   & 0.75   & 0.31   &[3]     &200	&2.13	 &(a)	  &$<$ 0.25 \\
\object{HD\, 22049}	&5073	 &4.43    &1.05    &-0.13   & 0.77   & --   & avg      &--	&2.13	 &(a)	  &$<$ 0.25 \\
\object{HD\, 23249}	&5074	 &3.77    &1.08    &0.13    &$<$ 0.15  &--      &[5]     &80	&1.01	 &(a)	  & 1.24 \\
\object{HD\, 23484}	&5176	 &4.41    &1.03    &0.06    &$<$ 0.70  &--      &[3]     &140	&2.40	 &(a)	  &$<$ 0.40 \\
\object{HD\, 26965}	&5126	 &4.51    &0.60    &-0.31   & 0.76   & 0.13   &[5]     &55	&0.77	 &(a)	  &$<$ 0.17 \\
\object{HD\, 27442}	&4825	 &3.55    &1.18    &0.39    &$<$ 0.30  &--      &[3]     &110	&$\sim$0 &(a)	  &$<$-0.47 \\
\object{HD\, 30495}	&5868	 &4.55    &1.24    &0.02    & 1.16   & 0.11   &[3]     &140	&3.04	 &(a)	  & 2.44  \\
\object{HD\, 36435}	&5479	 &4.61    &1.12    &0.00    & 0.99   & 0.12   &[3]     &210	&4.58	 &(a)	  & 1.67  \\
\object{HD\, 38529}	&5674	 &3.94    &1.38    &0.40    &$<$-0.10  &--      &[2]     &60	&--	 &--	  &$<$ 0.61 \\
\object{HD\, 38858}	&5752	 &4.53    &1.26    &-0.23   & 1.02   & 0.11   &[3]     &150	&0.99	 &(a)	  & 1.64  \\
\object{HD\, 43162}	&5633	 &4.48    &1.24    &-0.01   & 1.08   & 0.11   &[3]     &160	&5.49	 &(a)	  & 2.34  \\
\object{HD\, 43834}	&5594	 &4.41    &1.05    &0.10    & 0.94   & 0.11   &[3]     &220	&1.44	 &(a)	  & 2.30  \\
\object{HD\, 46375}	&5268	 &4.41    &0.97    &0.20    &$<$ 0.80  &--      &[3]     &90	&--	 &--	  &$<$-0.02 \\
\object{HD\, 52265}	&6103	 &4.28    &1.36    &0.23    & 1.25   & 0.11   &[1]     &120	&3.95	 &(a)	  & 2.88  \\
\object{HD\, 69830}	&5410	 &4.38    &0.89    &-0.03   & 0.79   & 0.11   &[3]     &100	&0.75	 &(a)	  &$<$ 0.47 \\
\object{HD\, 72673}	&5242	 &4.50    &0.69    &-0.37   & 0.70   & 0.13   &[3]     &180	&1.19	 &(a)	  &$<$ 0.48 \\
\object{HD\, 74576}	&5000	 &4.55    &1.07    &-0.03   & 0.70   & 0.31   &[3]     &120	&3.56	 &(a)	  & 1.72  \\
\object{HD\, 75289}	&6143	 &4.42    &1.53    &0.28    & 1.38   & 0.10   &[2]     &30	&3.81	 &(a)	  & 2.85  \\
\object{HD\, 75289}	&6143	 &4.42    &1.53    &0.28    & 1.33   & 0.12   &[1]     &110	&3.81	 &(a)	  & 2.85  \\
\object{HD\, 75289}	&6143	 &4.42    &1.53    &0.28    & 1.36   &--      &avg     &110	&3.81	 &(a)	  & 2.85  \\
\object{HD\, 76151}	&5803	 &4.50    &1.02    &0.14    & 1.02   & 0.11   &[3]     &110	&1.02	 &(a)	  & 1.88  \\
\hline
\end{tabular}
\\ $^\dagger$ The instruments used to obtain the spectra were: [1] UVES(A); [2] UVES(B); [3] UVES(C); [4] IACUB(A); [5] IACUB(B); [6] UES
\newline
$^{\dagger\dagger}$ The sources of the $v\,\sin{i}$ are: (a) CORALIE \citep[][]{San02b}; (b) \citet[][]{Gon97}; (c) \citet[][]{Gon98}; (d) \citet[][]{Nae04}; (e) \citet[][]{Sod82}
\newline
$^{\star}$ The solar Be abundance was computed using the Kurucz Solar Atlas \citep[][]{Kur84}; the solar Li abundance 
was taken from \citet[][]{Gre98}.
\label{table1}
\end{table*}

\begin{table*}
\caption[]{Derived Be abundances (HD numbers from 80\,000 on).}
\begin{tabular}{lcccrrcccccr}
\hline
Star     & T$_{\mathrm{eff}}$ & $\log{g}_{spec}$ & $\xi_{\mathrm{t}}$ & \multicolumn{1}{c}{[Fe/H]} & $\log{N(Be)}$ & $\sigma(Be)$ & Instr.$^\dagger$ & S/N & $v\,\sin{i}$ & source$^{\dagger\dagger}$ & $\log{N(Li)}$\\
         & [K]                &  [cm\,s$^{-2}$]  &  [km\,s$^{-1}$]    &                            &               &              &        &     & [km\,s$^{-1}$]&       &              \\
\hline
\object{HD\, 82943}	&6016	 &4.46    &1.13    &0.30    & 1.37   & 0.17   &[4]     &20	&1.65	 &(a)	  & 2.51  \\
\object{HD\, 82943}	&6016	 &4.46    &1.13    &0.30    & 1.27   & 0.12   &[2]     &35	&1.65	 &(a)	  & 2.51  \\
\object{HD\, 82943}	&6016	 &4.46    &1.13    &0.30    & 1.27   & 0.12   &[1]     &140	&1.65	 &(a)	  & 2.51  \\
\object{HD\, 82943}$^{\star\star}$	&6016	 &4.46    &1.13    &0.30    & 1.27   &--      &avg     &140	&1.65	 &(a)	  & 2.51  \\
\object{HD\, 83443}	&5454	 &4.33    &1.08    &0.35    &$<$ 0.70  &--      &[3]     &100	&1.38	 &(a)	  &$<$ 0.52 \\
\object{HD\, 84117}	&6167	 &4.35    &1.42    &-0.03   & 1.11   & 0.11   &[3]     &160	&4.85	 &(a)	  & 2.64  \\
\object{HD\, 92788}	&5821	 &4.45    &1.16    &0.32    & 1.19   & 0.11   &[2]     &40	&1.78	 &(a)	  & 1.34  \\
\object{HD\, 95128}	&5954	 &4.44    &1.30    &0.06    & 1.23   & 0.11   &[4]     &100	&2.1	 &(c)	  & 1.83  \\
\object{HD\,108147}	&6248	 &4.49    &1.35    &0.20    & 0.99   & 0.10   &[2]     &60	&5.34	 &(a)	  & 2.33  \\
\object{HD\,114762}	&5884	 &4.22    &1.31    &-0.70   & 0.82   & 0.11   &[4]     &65	&1.5	 &(c)	  & 2.20  \\
\object{HD\,117176}	&5560	 &4.07    &1.18    &-0.06   & 0.86   & 0.13   &[4]     &70	&0.5	 &(c)	  & 1.88  \\
\object{HD\,120136}	&6339	 &4.19    &1.70    &0.23    &$<$ 0.25  &--      &[6]     &90	&14.5	 &(b)	  &--	  \\
\object{HD\,121504}	&6075	 &4.64    &1.31    &0.16    & 1.33   & 0.11   &[2]     &45	&2.56	 &(a)	  & 2.65  \\
\object{HD\,130322}	&5392	 &4.48    &0.85    &0.03    & 0.95   & 0.13   &[4]     &35	&1.47	 &(a)	  &$<$ 0.13 \\
\object{HD\,134987}	&5776	 &4.36    &1.09    &0.30    & 1.22   & 0.11   &[2]     &60	&2.22	 &(a)	  &$<$ 0.74 \\
\object{HD\,143761}	&5853	 &4.41    &1.35    &-0.21   & 1.11   & 0.12   &[6]     &120	&1.5	 &(c)	  & 1.46  \\
\object{HD\,145675}	&5311	 &4.42    &0.92    &0.43    &$<$ 0.65  &--      &[4]     &65	&1	 &(d)	  &$<$ 0.03 \\
\object{HD\,168443}	&5617	 &4.22    &1.21    &0.06    & 1.11   & 0.13   &[4]     &55	&1.68	 &(a)	  &$<$ 0.78 \\
\object{HD\,169830}	&6299	 &4.10    &1.42    &0.21    &$<$-0.40  &--      &[3]     &130	&3.35	 &(a)	  &$<$ 1.16 \\
\object{HD\,179949}	&6260	 &4.43    &1.41    &0.22    & 1.08   & 0.10   &[3]     &100	&6.10	 &(a)	  & 2.65  \\
\object{HD\,187123}	&5845	 &4.42    &1.10    &0.13    & 1.08   & 0.12   &[4]     &55	&1.73	 &(d)	  & 1.21  \\
\object{HD\,189567}	&5765	 &4.52    &1.22    &-0.23   & 1.06   & 0.10   &[3]     &160	&1.29	 &(a)	  &$<$ 0.82 \\
\object{HD\,192263}	&4947	 &4.51    &0.86    &-0.02   &$<$ 0.90  &--      &[3]     &60	&2.02	 &(a)	  &$<$-0.39 \\
\object{HD\,192310}	&5069	 &4.38    &0.79    &-0.01   &$<$ 0.60  &--      &[3]     &180	&0.85	 &(a)	  &$<$ 0.20 \\
\object{HD\,195019}	&5842	 &4.32    &1.27    &0.08    & 1.15   & 0.12   &[4]     &50	&1.73	 &(a)	  & 1.47  \\
\object{HD\,202206}	&5752	 &4.50    &1.01    &0.35    & 1.04   & 0.11   &[3]     &130	&2.44	 &(a)	  & 1.04  \\
\object{HD\,209458}	&6117	 &4.48    &1.40    &0.02    & 1.24   & 0.11   &[3]     &150	&3.65	 &(a)	  & 2.70  \\
\object{HD\,210277}	&5532	 &4.29    &1.04    &0.19    & 0.91   & 0.13   &[1]     &110	&1.39	 &(a)	  &$<$ 0.30 \\
\object{HD\,211415}	&5890	 &4.51    &1.12    &-0.17   & 1.12   & 0.10   &[3]     &190	&1.84	 &(a)	  & 1.92  \\
\object{HD\,217014}	&5804	 &4.42    &1.20    &0.20    & 1.02   & 0.12   &[6]     &100	&2.1	 &(c)	  & 1.30  \\
\object{HD\,217107}	&5646	 &4.31    &1.06    &0.37    & 0.96   & 0.13   &[1]     &120	&1.37	 &(a)	  &$<$ 0.40 \\
\object{HD\,222335}	&5260	 &4.45    &0.92    &-0.16   & 0.66   & 0.22   &[1]     &110	&1.25	 &(a)	  &$<$ 0.31 \\
\object{HD\,222582}	&5843	 &4.45    &1.03    &0.05    & 1.14   & 0.11   &[3]     &125	&1.75	 &(a)	  &$<$ 0.59 \\
\hline
\end{tabular}
\\ $^\dagger$ The instruments used to obtain the spectra were: [1] UVES(A); [2] UVES(B); [3] UVES(C); [4] IACUB(A); [5] IACUB(B); [6] UES
\newline
$^{\dagger\dagger}$ The sources of the $v\,\sin{i}$ are: (a) CORALIE \citep[][]{San02b}; (b) \citet[][]{Gon97}; (c) \citet[][]{Gon98}; (d) \citet[][]{Nae04}
\newline
$^{\star\star}$ Given the lower S/N of the IACUB spectrum, only the UVES spectra were considered.
\label{table2}
\end{table*}

Most studies of light elements in solar-type stars have been based on
Li abundances. This is mostly because Li features are
easier to measure from high-resolution optical spectra (e.g. the Li6708\AA\ line). 
On the other hand, the only available Be features (\ion{Be}{ii}) are located 
near the atmospheric near-UV cutoff. Using ground-based telescopes to measure Be 
abundances is thus not a simple task. The situation for B is even worst, and 
only with instruments in space is it possible to obtain the abundances for 
this element. It is important, however, to complement the Li studies with 
analysis of Be and (if possible) B in solar-type stars. These three elements 
are destroyed by (p,$\alpha$)-reactions at different temperatures (about 2.5, 
3.5, and 5.0$\times$10$^{6}$\,K, for Li, Be and B, respectively). Their abundances 
can thus help us to probe different regions (depths) inside a solar-type star.

\begin{figure*}[t]
\psfig{width=\hsize,file=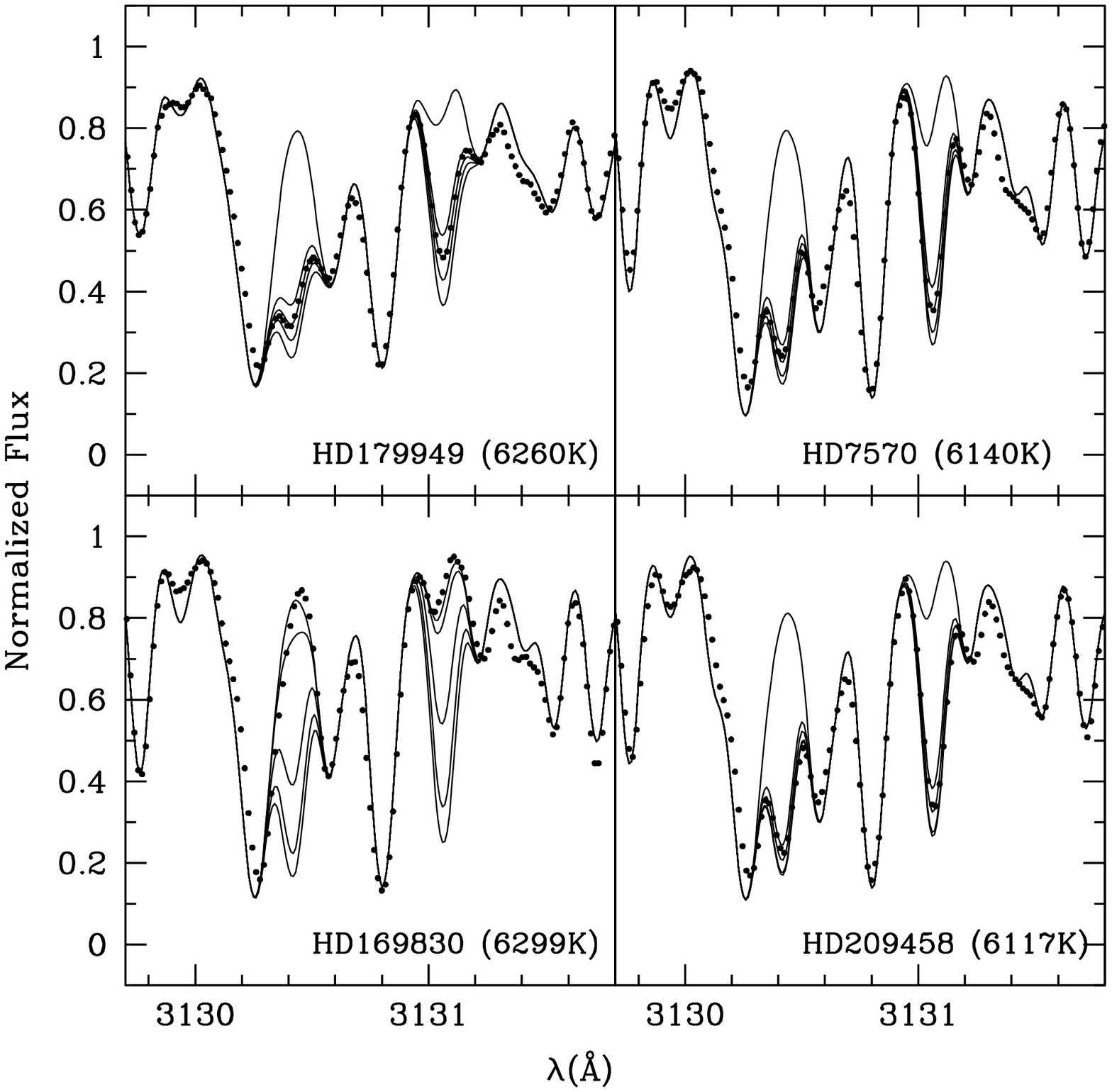}
\caption[]{Spectral syntheses (continuous lines) and observed spectra (points) in 
the \ion{Be}{ii} line region for 4 of the hottest dwarfs in our sample. In 
all panels the upper and lower syntheses were done with a $\log{N(Be)}$ 
of 1.42 (meteoritic) and $-$10.0 (essentially no Be), respectively. The three
intermediate syntheses for all stars except \object{HD\,169830} correspond to the optimal fit 
and to fits with abundance variations of $\pm$0.15\,dex. For \object{HD\,169830}, these
correspond to abundances of 1.15, 0.65, and $-$0.35\,dex. Stellar effective temperatures
are also shown.}
\label{fighot}
\end{figure*}

Beryllium is produced by spallation reactions in the interstellar medium, 
while it is burned in the hot stellar furnaces 
\citep[e.g.][]{Ree94,Ram97}. Since it is burned at much higher temperatures than Li, Be is depleted at lower rates, and thus we can expect to measure Be 
abundances in stars which have no detectable Li in their atmospheres
(like intermediate-age late G- or K-type dwarfs). 
In fact, for about 50\% of the known planet host stars no Li was detected 
\citep[e.g.][]{Isr04}. Furthermore, Li studies have shown the presence
of a significant scatter not only among field stars \citep[][]{Pas94}, but also among 
coeval stars in open clusters \citep[][]{Sod93,Gar94,Pas97,Ran98,Jon99}. 
More specifically, a spread is seen among K-type stars
in young clusters; this dispersion appears to be related to rotation
or rotational history. A dispersion is also seen among solar-type stars
in the solar age cluster M67; the reasons for this dispersion are still
unknown. Also note that the dispersion is not observed for other old
clusters \citep[][]{Ran03}. The presence of a dispersion
at a given age can indeed complicate any study.

In a companion paper to this -- \citet[][]{papera}, hereafter Paper\,A -- we present 
a study of Be in a set of field stars known to harbor planetary mass companions, 
as well as in a comparison set of ``single'' stars. The main result of 
that paper is that the differences between the two groups of stars 
are very small, if any. In the current paper we use the same sample 
studied in Paper\,A to explore the Be depletion throughout the main-sequence
and sub-giant branch. The results reveal some interesting and unexpected features. 
First, we confirm former results \citep[][]{San02} supporting that lower-temperature 
dwarfs burn Be more easily than their higher temperature counterparts. 
Be abundances seem to follow a trend similar to the one found for Li. 
We further show that there seems to exist a Be-dip for some solar-temperature dwarfs, 
and report on some sub-giants that have no measurable Be 
in their atmospheres, though one of them has detectable amounts of Li. 
Finally, Li and Be abundances for the same stars
are compared, giving us the possibility to determine the temperature at which
the onset of Li and Be depletion occurs.

\begin{figure*}[t]
\psfig{width=\hsize,file=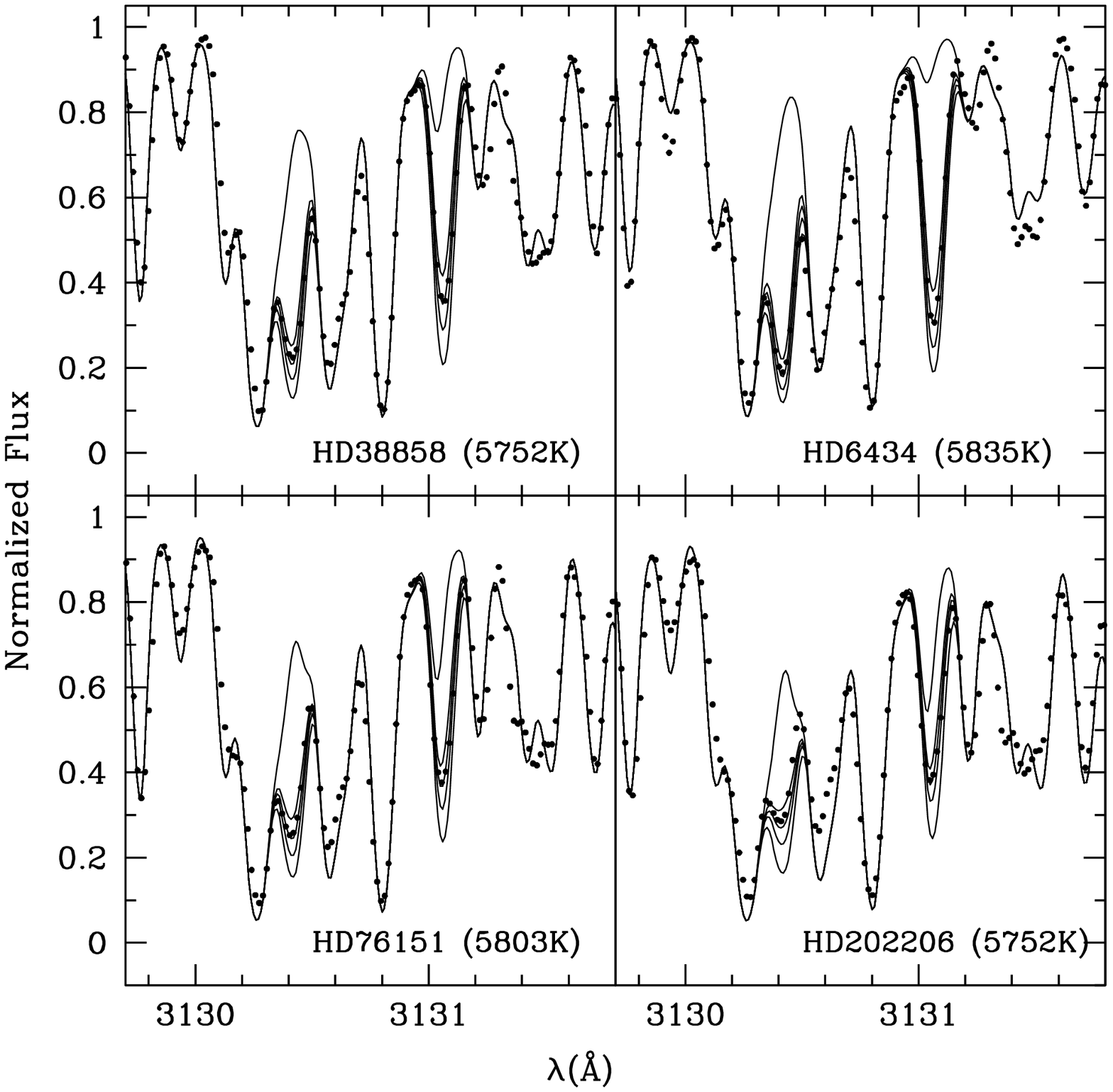}
\caption[]{Spectral syntheses (continuous lines) and observed spectra (points) in 
the \ion{Be}{ii} line region for 4 solar-temperature dwarfs in our sample. In 
all panels, the upper and lower syntheses were done with a $\log{N(Be)}$ 
of 1.42 (meteoritic) and $-$10.0 (essentially no Be), respectively. For all stars
the three intermediate syntheses correspond to the optimal fit 
and to fits with abundance variations of $\pm$0.15\,dex. }
\label{figsolar}
\end{figure*}

\begin{figure*}[t]
\psfig{width=\hsize,file=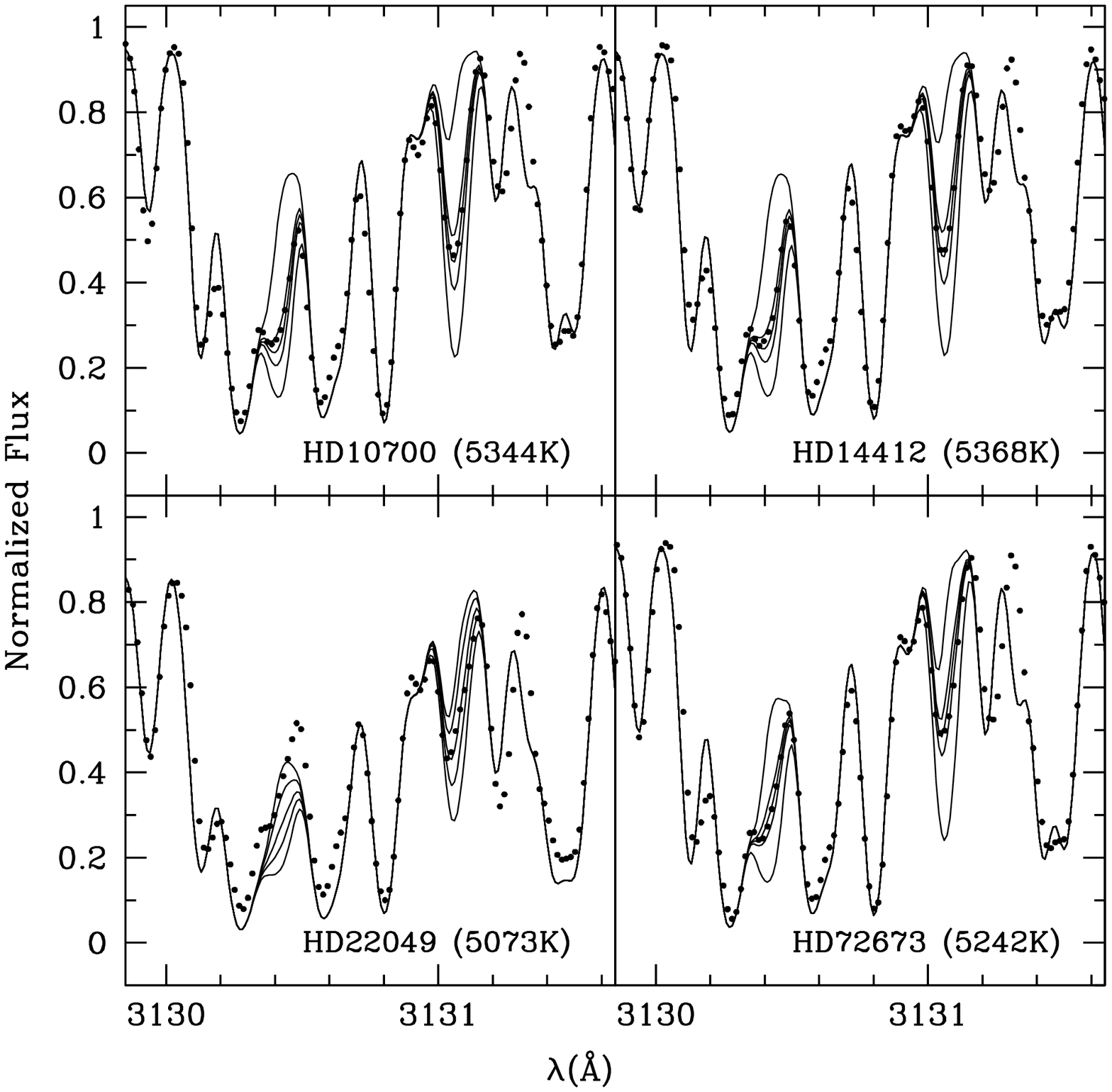}
\caption[]{Spectral syntheses (continuous lines) and observed spectra (points) in 
the \ion{Be}{ii} line region for 4 of the coolest dwarfs in our sample. In 
all panels, the upper and lower syntheses were done with a $\log{N(Be)}$ 
of 1.42 (meteoritic) and $-$10.0 (essentially no Be), respectively. For all stars
expect \object{HD\,22049}, the three intermediate syntheses correspond to the optimal fit 
and to fits with abundance variations of $\pm$0.15\,dex. For this star,
the three intermediate fits were done using Be abundances of 1.05, 0.75 (best fit), 
and 0.25\,dex.}
\label{figcool}
\end{figure*}

\section{The data, stellar parameters, and chemical analysis}
\label{sec:data}

The data used in this paper has already been fully described in Paper\,A. 
We refer to this for more details on the spectrographs/telescopes
used, S/N and resolution of the spectra, and for data-reduction details. 

Except for \object{BD\,$-$10\,3166} \citep[][]{Gon01}, all the stellar parameters were 
taken from the fully spectroscopic uniform study of \citet[][]{San04b},
or derived using the same techique. Given that these 
parameters have been obtained using the same technique for all our stars, this gives us
an important guarantee of uniformity in the analysis.

The Be abundances were then derived by fitting the spectral region around the \ion{Be}{ii} 
lines at 3130.420 and 3131.065\,\AA. In practice, only the latter line was used to derive the Be abundances, 
given the severe line-blending in the region around 3130.42\AA; the bluest line was thus
only used to check the consistency of the fit. The analysis was done assuming LTE,
using the 2002 version of the code MOOG \citep{Sne73}, and a grid of \citet{Kur93} 
ATLAS9 atmospheres. The line list used was described in \citet{Gar98}.
The final precision of the abundance results is of the order of 0.15\,dex. 
The Be abundance we have obtained from an analysis of the Kurucz Solar Atlas \citep[][]{Kur84} is 1.10, 
close to the value of 1.15 obtained by \citet[][]{Chm75}. Again, for more details we point 
the reader to Paper\,A. 

Finally, the Li abundances are a revised version of the values derived in
\citet[][]{Isr04}, and the values of $v\,\sin{i}$ were taken, in most of the cases, from
a calibration of the CORALIE cross-correlation function \citep[see appendix in][]{San02b}. 
In the rest of this paper, the solar Li abundance was taken 
from \citet[][]{Gre98}, $\log{N(Li)}$=1.10.
  
In Tables\,\ref{table1} and \ref{table2} we present the final Be abundances, stellar parameters,
$v\,\sin{i}$ values, and Li abundances.

\section{Be as a function of T$_{\mathrm{eff}}$}
\label{sec:beteff}

\begin{figure}[t]
\psfig{width=\hsize,file=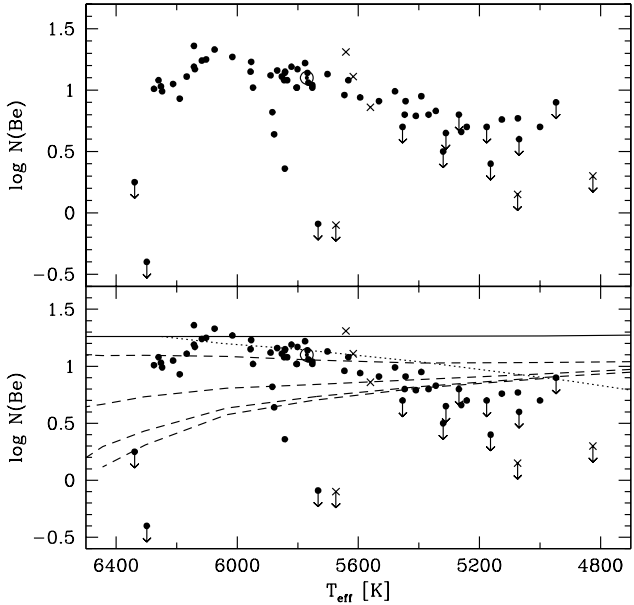}
\caption[]{{\it Upper panel}: Derived Be abundances as a function of T$_{\mathrm{eff}}$.
Dots represent dwarfs, while crosses denote sub-giants. The Sun is
denoted by the usual symbol. {\it Lower panel}: Same as above, but superposing the Be depletion
models of \citet[][]{Pin90} (Case A) for solar metallicity and an age of 1.7\,Gyr (dashed lines). 
From top to bottom, the lines represent a standard model (solid line), and 4 models with different initial angular momentum \citep[Tables\,3 to 6 of ][]{Pin90}. The solid line represents the initial Be abundance of 1.26 (see text). The dotted line represents the Be depletion isochrone for 4.5\,Gyr taken from
the models including mixing by internal waves of \citet[][]{Mon00}.}
\label{figbeteff}
\end{figure}

In Fig.\,\ref{figbeteff} (upper panel) we plot the derived Be abundances as 
a function of effective temperature for our program stars. In the plot, 
the circles denote dwarfs, while sub-giants are denoted with crosses.
The Sun is denoted by the usual solar symbol, assuming a Be abundance of 
1.10 (see Sect.\,\ref{sec:data}). 

\begin{figure}[t]
\psfig{width=\hsize,file=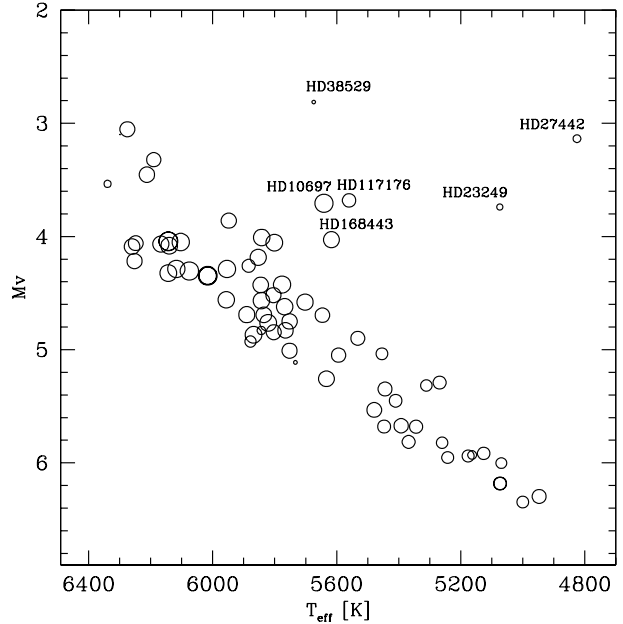}
\caption[]{HR diagram for the stars studied in this paper. The size of
the circles is proportional to the Be abundance. Six of the stars (labeled)
fall considerably outside the main-sequence, and are thus probably sub-giants.
Absolute magnitudes were computed from Hipparcos parallaxes and V magnitudes \citep[][]{ESA97}.
For clarity, no difference is made between stars with Be detections and
with only an upper limit.}
\label{fig_hr}
\end{figure}

Sub-giants were defined as those stars that fall considerably above the 
Main-Sequence in the HR diagram of Fig.\,\ref{fig_hr}. There are
6 of those, namely \object{HD\,10647}, \object{HD\,23249}, \object{HD\,27442},
\object{HD\,38529}, \object{HD\,117176}, and \object{HD\,168443}. Except for this
latter star, all the others have surface gravity values below 4.1\,dex (\object{HD\,168443}
has a $\log{g}$=4.22), attesting to their evolutionary status. 

The Be abundances for the 
sub-giants should be considered separately, since their 
effective temperatures may have changed considerably from the value they had 
when the stars were on the main-sequence; extra mixing may already have occurred, 
causing dilution and/or depletion of their light-element contents. 

A look at Fig.\,\ref{figbeteff} (upper panel) shows that the Be abundances have a 
maximum near 6100\,K. 
Very little, if any, Be depletion has occurred for stars of this effective temperature. 
A similar maximum for the Li abundances is found at about the same effective 
temperature (see e.g. Fig.\,\ref{figbeteffgap}). For higher temperatures, 
the values of $\log{N(Be)}$ decrease with increasing temperature, forming 
the well known Be gap for F stars \citep[e.g.][]{Boe02}, a feature that has a 
counterpart for Li. For temperature values lower than $\sim$6100\,K, and except for
a few outliers, the Be abundances also show a clear but less steep decreasing trend 
with decreasing temperature. This trend was already mentioned and 
discussed in \citet[][]{San02}, and is similar to that typically found for Li.

As discussed in Paper\,A, the maximum seen near 6100\,K may be 
attributed e.g. to Galactic chemical evolution effects, since most of the stars in the 
temperature interval between 6000 and 6200\,K are particularly metal-rich,
and Galactic Be abundances are known to increase with the 
metallicity \citep[][]{Reb88,Boe93,Mol97,Boe99}.
If so, and excluding these objects, Be abundances might form a plateau for temperatures 
between $\sim$6000 and 5600\,K (see also Sect.\,\ref{sec:libe}), decreasing 
with decreasing temperature 
for T$_{\mathrm{eff}}$ below $\sim$5600\,K. The cause for the observed trend
for the lower temperature dwarfs in our sample can hardly be attributed
to any metallicity effect, however, since some very metal-rich stars are 
present at these low temperatures.

Interestingly, this latter result is not confirmed by other studies of Be abundances in open 
clusters F and G dwarfs \citep[e.g.][]{Boe02,Boe03a,Boe03b}. It should be mentioned, however, that 
in their studies these authors have not analyzed stars with temperatures 
much below 5500\,K. Furthermore, the stars studied by \citet[][]{Boe02} and \citet{Boe03a,Boe03b} 
are considerably younger than the ``average'' field stars in our sample. The fact that 
we do observe a decrease of the Be abundances as a function of decreasing temperature can 
be seen as evidence that Be is burned during the main-sequence evolution of these stars.

In the lower panel of Fig.\,\ref{figbeteff} we present the same plot but
now we have superposed on the Be measurements a set of Yale beryllium 
depletion isochrones from \citet[][]{Pin90} -- dashed lines -- for solar metallicity and an 
age of 1.7\,Gyr (to our knowledge there are no isochrones available for higher ages).
Given that it is not possible to know what was the initial Be abundance 
for all the targets, we will assume an initial $\log{N(Be)}$ of 1.26 for all 
stars, i.e., between Solar (1.10 -- see Sect.\,\ref{sec:data}) and meteoritic \citep[1.42 - ][]{And89}. 

For temperatures above roughly 5600\,K, the \citet[][]{Pin90} models seem to follow 
reasonably well the observed trends. This confirms the result 
of \citet[][]{Ste97}, who have already shown that these
models (improved Yale models by \citet[][]{Del93} do not 
show a clear difference) are quite satisfactory for stars with 5600\,K$\leq$T$_{\mathrm{eff}}$$\leq$6500\,K.
However, it is also clear from the figure that there seems to be a clear 
discrepancy between the model predictions and the observed abundances for 
lower temperature stars. As already noted in \citet[][]{San02}, while for
these objects we 
observe a decrease in the Be abundance as a function of decreasing 
temperature, the model predictions go in the opposite sense.  

There are several possibilities for explaining this observed inconsistency. 
One way would be to say that the derived abundances have some bias as a function 
of effective temperature. However, and as we can see from the plots of Fig.\,\ref{figcool}, 
our synthetic spectra seem to fit quite well the whole Be-line region for the lower 
temperature stars in our samples. Although we consider it quite unlikely that 
errors in the spectroscopic analysis are causing the whole observed trend,
we should mention that the \ion{Be}{ii} line region is very crowded. \citet[][]{Gar95} 
studied in detail the sensitivity of the observed feature at $\lambda$\,3131\,\AA\ to 
Be abundance for low-mass stars. For effective temperatures below $\sim$5100\,K the 
feature starts to be dominated by the contribution of another element (likely \ion{Mn}{i}). 
This makes it difficult to determine accurate Be abundances for the coolest
stars of our sample.

As already pointed out in Paper\,A, the usefulness of the \ion{Be}{ii} line for 
deriving Be abundances will thus 
decrease with temperature to a point where it will be useless. However, the 
exact temperature at which this occurs will depend on the resolution and signal-to-noise 
ratio of the observed spectra. Even if the observed feature is not dominated by the 
Be transition, it is possible to obtain a reliable Be measurement by fitting synthetic
spectra with different Be abundances, as long as the S/N of the data is high enough.

In our case, the increasing dispersion observed for Be abundances of stars with
temperature below $\sim$5400\,K possibly reflects the increasing difficulty in deriving 
the Be abundances for the lowest temperature stars, although a real ``physical'' reason cannot 
be completely excluded. However, it should be mentioned that
the decreasing trend of Be abundances as a function of decreasing T$_{\mathrm{eff}}$ starts already 
for temperature values only $\sim$150\,K below the solar temperature (for which the line list was built). 
This makes very unlikely that the observed trend may be attributed to errors in the spectroscopic
analysis only.

In this sense, one other possible source of error could arise from the presence of 
a ``missing near-UV opacity'', as discussed by \citet[][]{Bal98}, something that we 
have not considered. However, the need to include this extra opacity is not established. \citet[][]{All00}, 
for example, have shown that it is possible to make a good fit of the solar spectrum without taking into 
account this extra-opacity. Furthermore, the problem discussed by \citet[][]{Bal98} refers to the solar 
case, and has not been discussed for other temperature regimes.

Another possibility for explaining the observed trend would be to consider
that this has something to do with the presence of planets. Indeed, about one half 
of the stars in our sample are planet-hosts. If stellar pollution is very important among 
these objects, we could probably expect earlier-type dwarfs to present higher Be abundances, given
that they have shallower convective envelopes. However, and as discussed in Paper\,A,
there does not seem to be a large difference in the Be content of stars with and
without known giant planets. We thus have no reason to believe that the presence of planets
is causing the observed inconsistency. Curiously, \citet[][]{Isr04} have shown that
in the temperature interval between 5600 and 5850\,K planet hosts are Li deficient
when compared with field dwarfs. However, both samples do present the same Be trend as a function
of T$_{\mathrm{eff}}$ (see Paper\,A). In other words, the difference noticed for Li is aparentely not found 
for Be. We should mention, however, that the relatively low number of points may
preclude a firmer conclusion.

\citet[][]{Mur01} have suggested that the stars in the solar neighborhood have 
accreted an average of $\sim$0.5\,M$_{\sun}$, 
a value compatible with the upper limit found for the Hyades cluster by \citet[][]{Qui02}. 
This small amount of material, which would pollute more efficiently the hottest stars in our sample
(these have shallower convective envelopes) is, however, probably not enough to explain the observed 
increasing trend.

It has been shown by \citet[][]{Swe94} that even a small change in the
oxygen abundance may have a strong effect on the Li depletion rate
in solar-type stars. This effect is due to the change of the depth of the convective 
envelope induced by the change of the global opacity. Although we do not
have oxygen abundances for our targets, they are spread over a wide range 
of metallicities. It is thus not very likely that the observed inconsistency 
is due to such an effect.

Interestingly, the models of \citet[][]{Mon00}, which include mixing by internal waves 
(dotted line\footnote{The curve was taken directly from their Fig.\,4.} in Figure\,\ref{figbeteff}), 
could solve at least part of the problem, since they do predict an increasing Be
depletion as a function of decreasing temperature (see also their Fig.\,4).
While this model still overestimates the Be abundances for the cooler stars in our
sample, it fits our data much better than the \citet[][]{Pin90} models do.

\section{A Be gap for solar-temperature dwarfs}
\label{sec:gap}

\begin{figure}[t]
\psfig{width=\hsize,file=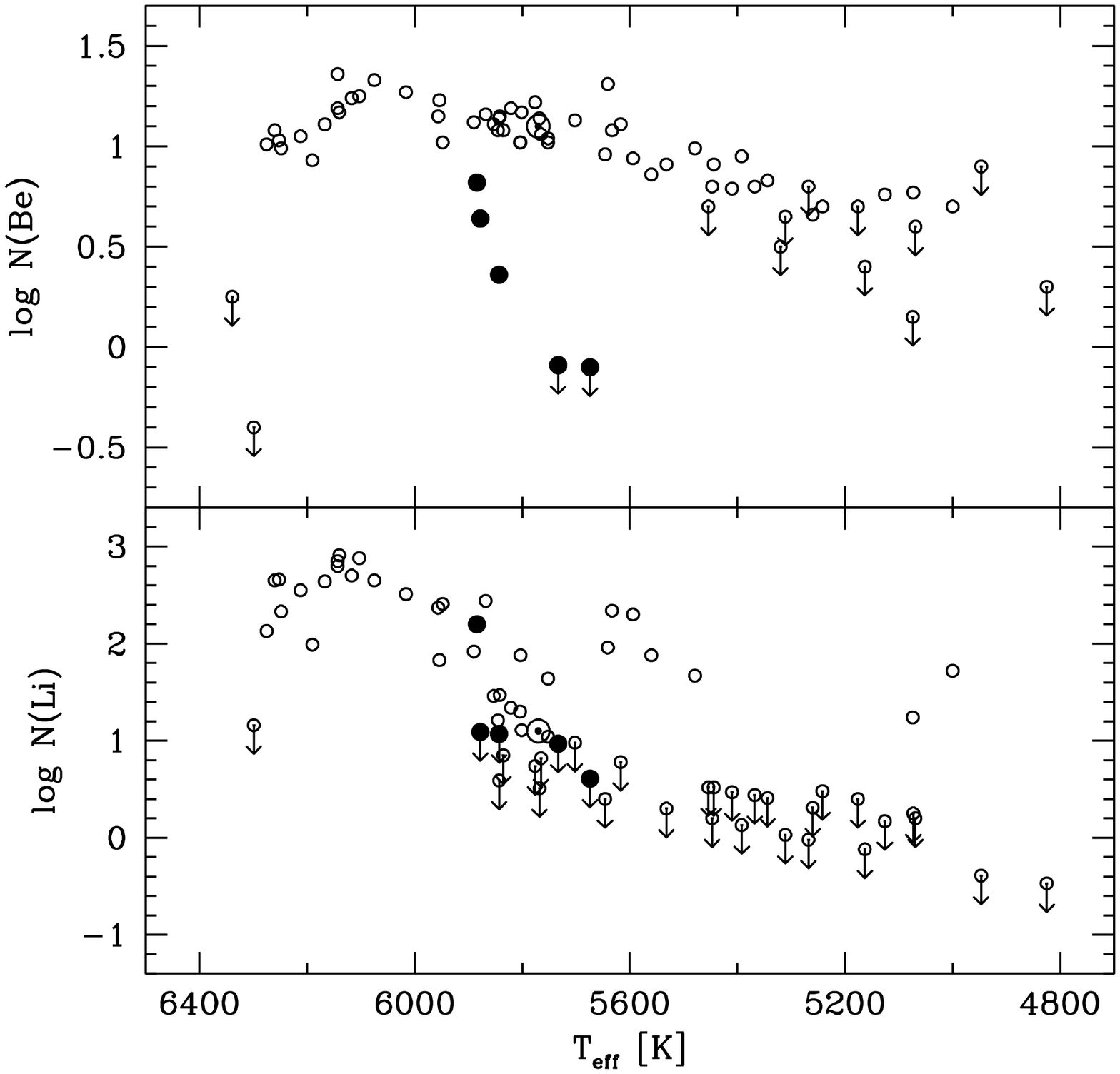}
\caption[]{{\it Upper panel}: Be abundances as a function of effective temperature for
the stars in our sample. No distinction was made between dwarfs and sub-giants. 
The 5 stars (4 dwarfs and 1 sub-giant) with anomalous Be abundances are denoted by 
the filled circles (see text and Fig.\,\ref{figgap} for more details). {\it Lower panel}: same as above,
but for Li.}
\label{figbeteffgap}
\end{figure}

\begin{figure*}[t]
\psfig{width=\hsize,file=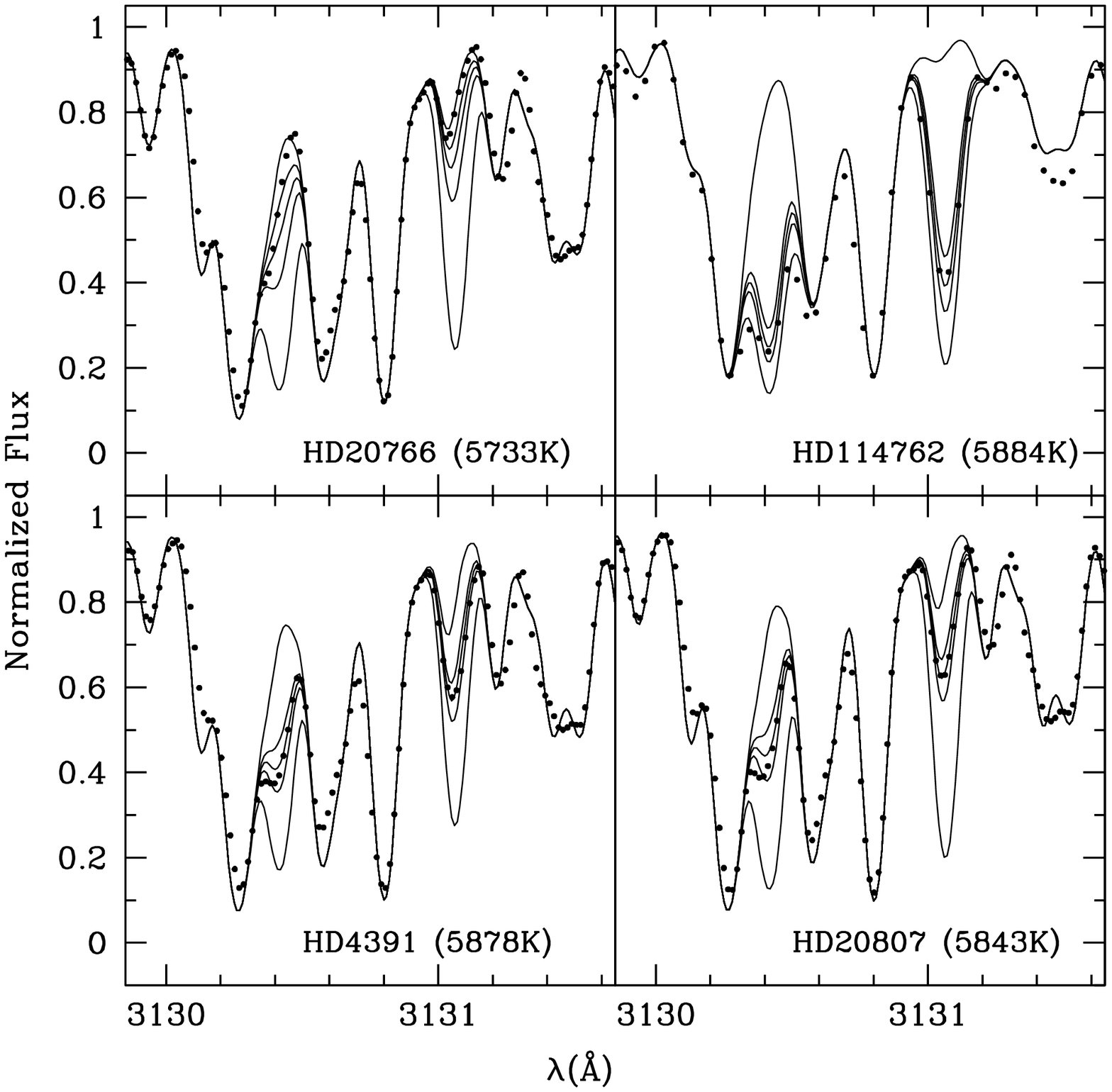}
\caption[]{Spectral syntheses (continuous lines) and observed spectra (points) in 
the \ion{Be}{ii} line region for four stars discussed in Sect.\,\ref{sec:gap}. In 
all panels, the upper and lower syntheses were done with a $\log{N(Be)}$ 
of 1.42 (meteoritic) and $-$10.0 (essentially no Be), respectively. The three
intermediate syntheses for all but \object{HD\,20766} correspond to the optimal fit 
and to fits with abundance variations of $\pm$0.15\,dex. For this particular star, 
the three intermediate syntheses were done with Be abundances of 0.50, 0.20, and -0.10.}
\label{figgap}
\end{figure*}

In Fig.\,\ref{figbeteffgap} (upper panel) we present the same plot of Fig.\,\ref{figbeteff}, 
but highlighting a small group of solar-temperature stars that seem to present abnormally 
low Be abundances, falling completely outside of the main trend. The spectral syntheses for 
four of these stars are presented in Fig.\,\ref{figgap}, and for the remaining in Fig.\,\ref{figgiant}. 
These objects, all in the temperature 
regime between roughly 5600\,K and 5900\,K, lie in some kind of Be-gap, 
similar to the one suggested to exist for Li in the same temperature 
regime \citep[][]{Pas94,Che01}.

To understand the situation it is important to check the nature of these
particular stars. Four of them (\object{HD\,4391}, \object{HD\,20766}, 
\object{HD\,20807}, and \object{HD\,114762}) are 
dwarfs, while \object{HD\,38529} is one of the most evolved stars in our sample,
being already in the sub-giant regime (see Tables\,\ref{table1} and \ref{table2}). This latter case
will be discussed in the next section. 

As mentioned in Paper\,A, \object{HD\,4391}, \object{HD\,20766}, and
\object{HD\,20807} are not known to harbor any planetary-mass 
companion. \object{HD\,114762} and \object{HD\,38529},
on the other hand, do have low mass companions in orbit.

We have estimated the ages of \object{HD\,4391}, \object{HD\,20766}, and 
\object{HD\,20807} from their chromospheric 
activity \citep[derived from \ion{Ca}{ii} H and K lines -- ][]{Vau78}. 
Values for the Mount-Wilson S index were taken from \citet[][]{Hen96} and from
an analysis of the CORALIE spectra, a process described in \citet[][]{San00}. 
Given the similarity of
the values, we have decided to take the average S$_{MW}$ value for each star
to compute the stellar age. This latter was derived from the calibration
presented in \citet[][]{Hen96} \citep[taken from][]{Don93}. 
The knowledge of the S index also gives us the possibility to access the
rotational period of the stars \citep[][]{Noy84}. The results are 
summarized in Table\,\ref{tabgap}.  

\begin{table}[t!]
\caption[]{Data for \object{HD\,4391}, \object{HD\,20766}, and \object{HD\,20807} (see text for more details)}
\begin{tabular}{lccccc}
\hline
Star               & S$_{\mathrm{MW}}$ & $B-V$$^c$ &   Age  & P$_{\mathrm{rot}}$ & $v\,\sin{i}$$^d$ \\
                   &                   &           &  (Gyr) &      (days)        &   (km\,s$^{-1}$) \\
\hline
\object{HD\,4391}  & 0.28$^a$/0.26$^b$ &  0.64 & 1.2       & 12    & 2.72 \\
\object{HD\,20766} & 0.23$^a$/0.26$^b$ &  0.64 & 1.6       & 14    & 1.98 \\
\object{HD\,20807} & 0.19$^a$/0.23$^b$ &  0.60 & 2.1       & 13    & 1.74 \\
\hline
\end{tabular}
\newline
$^a$ Taken from \citet[][]{Hen96}
\newline
$^b$ Derived using CORALIE spectra \citep[][]{San00}
\newline
$^c$ Taken from the Hipparcos catalogue \citep[][]{ESA97}
\newline
$^d$ Derived from the width of the CORALIE cross-correlation function \citep[][]{San02b}
\label{tabgap}
\end{table}

An inspection of Table\,\ref{tabgap} shows that all these stars seem to
have very similar ages, rotation periods, as well as projected rotational velocities.
Furthermore, and as can be seen from Table\,\ref{table1}, their metallicities are
all slightly below solar. Actually, these three stars
are not very different from many other objects listed in Tables\,\ref{table1} and \ref{table2}, occupying
the same temperature regime (including our Sun; good examples are \object{HD\,38858}
and \object{HD\,189567}\footnote{These two stars also have similar or lower chromospheric
activity levels, implying similar or higher ages.}) but having ``normal'' Be abundances. 

\object{HD\,4391}, \object{HD\,20766}, and \object{HD\,20807}
only have upper-limit estimates for the Li abundance, making it difficult
to compare the Li/Be ratios (see Fig.\,\ref{figbeteffgap}, lower panel). 

For \object{HD\,114762}, however, the Li abundance is quite high ($\log{N(Li)}$=2.20), 
which poses the problem of explaining how this star has become fairly depleted in Be 
(by a factor of $\sim$4, if we suppose that the original Be abundances was similar 
to the one found in meteorites) while keeping so much Li in its atmosphere 
(compared to the meteoritic value of 3.31, \citet[][]{And89}, it is depleted by a factor of 12). 
However, one must take into account that \object{HD\,114762} is the most deficient 
star in our sample. Due to the chemical evolution of Be in the galaxy, this might imply that
the original Be abundance was considerably lower than meteoritic for this
object \citep[e.g.][]{Reb88,Boe93,Mol97,Boe99} -- see also Garc\'{\i}a L\'opez et al. (in preparation).
Indeed, its Be abundance is perfectly compatible with the one found for stars of 
its metallicity \citep[e.g.][]{Boe99}.

In any case, the three remaining dwarfs (\object{HD\,4391}, \object{HD\,20766}, 
and \object{HD\,20807}) clearly show that there seems to be an unexpected Be-gap
for solar temperature stars, a region for which, at least in some cases,
Be depletion is particularly severe. Some process capable of significantly 
increasing the mixing in the stellar interior seems to have played some role,
and produced the observed scatter in the Be abundances.
This process may be related to the possibility that different pre-main-sequence 
rotational histories may exist for different stars of similar mass, age, 
and metallicity \citep[see e.g. ][]{Bar03}.

Finally, it is interesting to see that \object{HD\,20766} and \object{HD\,20807}, 
two stars of the known wide visual binary system $\zeta^1$ and $\zeta^2$ Ret (with 
similar radial velocities and proper motions), are the two most Be-poor stars among 
the four objects discussed above. This fact could hint that
a bona-fide explanation (although unlikely) for their very low Be abundances,
is an abnormally low initial Be content. Curiously though, these stars have previously 
been noticed to be somehow abnormal \citep[c.f.][]{Sil87,Roc02}.

\section{Sub-giants}
\label{sec:giants}

Six stars in our sample have a position considerably above the HR diagram (see Fig.\,\ref{fig_hr}). 
These objects are probably sub-giants, stars that are already evolving out of the
main sequence. During this process, the convective envelopes of the stars deepen, and
the light-element abundances can suffer considerable depletion/dilution. It is
thus wise to treat these cases separately from the other stars (dwarfs).

\begin{figure*}[t]
\psfig{width=\hsize,file=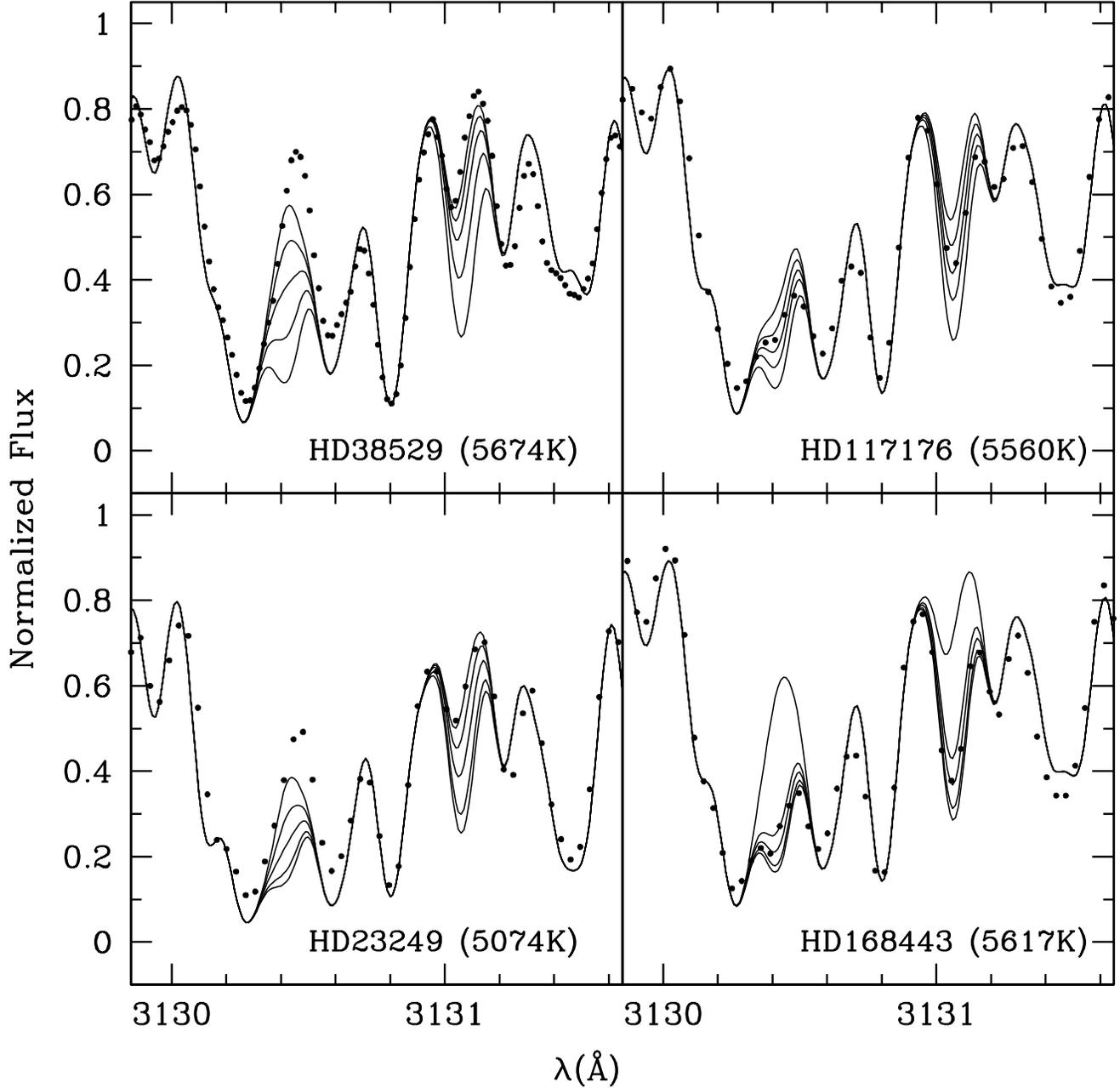}
\caption[]{Spectral syntheses (continuous lines) and observed spectra (points) in 
the \ion{Be}{ii} line region for four of the stars discussed in Sect.\,\ref{sec:giants}. In all panels, 
the upper and lower syntheses were done with a $\log{N(Be)}$ of 1.42 (meteoritic) and $-$10.0 
(essentially no Be), respectively. The three other spectral 
syntheses for \object{HD\,23249} correspond to Be abundances of 1.15, 0.65, and 0.15\,dex, while for \object{HD\,38529} 
these correspond abundances of 0.80, 0.35, and $-$0.10. For \object{HD\,117176} and
\object{HD\,168443} the three intermediate fits correspond to the optimal fit and to fits with abundance 
variations of $\pm$0.20\,dex. }
\label{figgiant}
\end{figure*}

\begin{figure}[t]
\psfig{width=\hsize,file=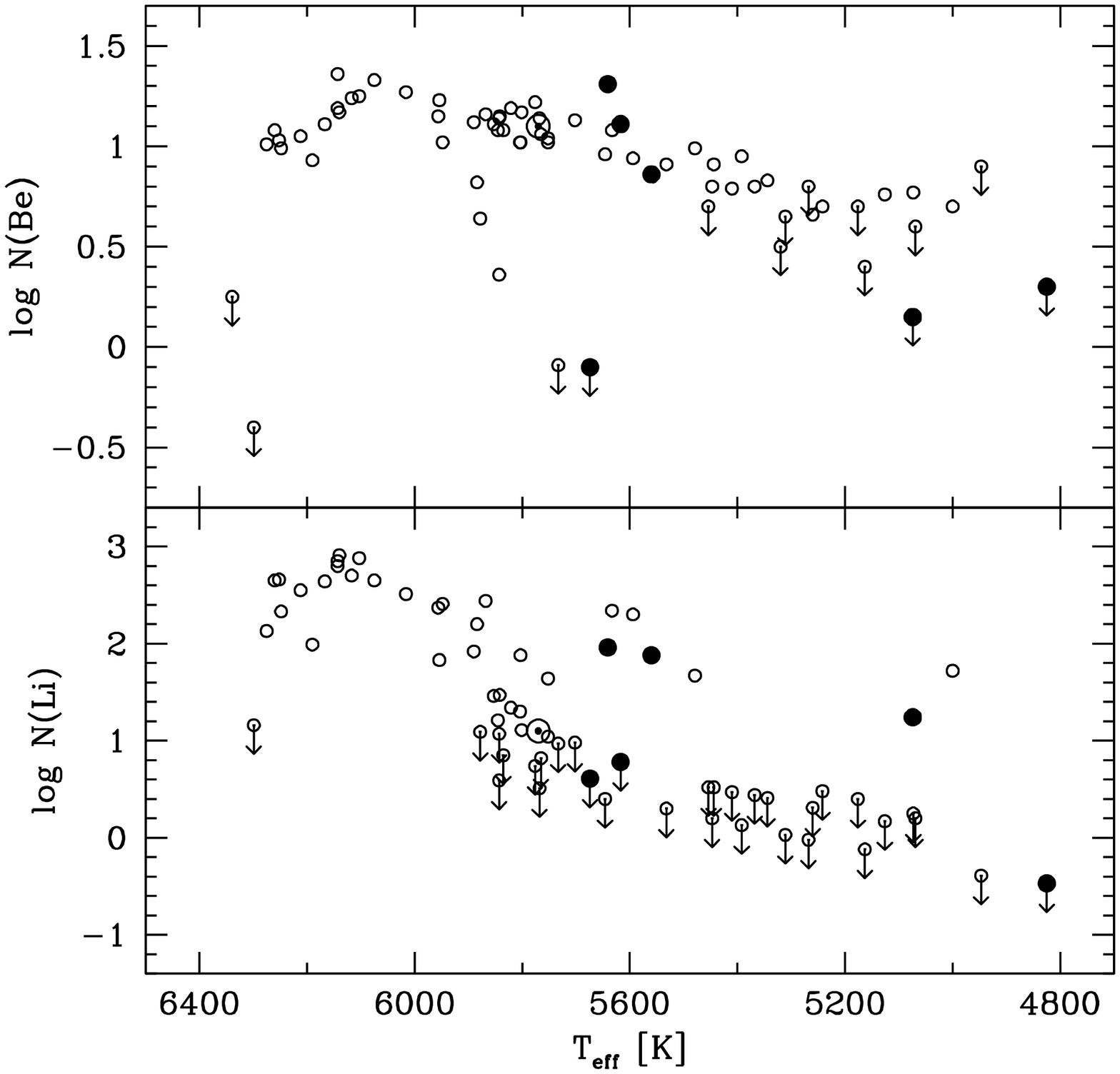}
\caption[]{{\it Upper panel}: Be abundances as a function of effective temperature for
the stars in our sample. The 6 sub-giant stars are 
denoted by the filled circles. {\it Lower panel}: same as above, but for Li.}
\label{figbeteffgiants}
\end{figure}

In Fig.\,\ref{figbeteffgiants} we present a plot of the Be and Li abundances as
a function of effective temperature, calling the attention to the position
of the sub-giants in the plot (filled circles) -- see also Fig\,\ref{figgiant}.
As a general comment, these stars do follow the general trend of Be vs. T$_{\mathrm{eff}}$
observed for dwarfs. Let us discuss these cases in more detail, separating these
sub-giants into four groups, as follows.\\

{\it \object{HD\,27442} and \object{HD\,38529}}. 
These two stars (both planet-hosts -- Paper\,A, with [Fe/H]$\geq$0.21), with very different temperatures, 
both have depleted their Li and Be abundances; only upper limits of these 
were derived, making it impossible to explore their Li/Be ratios. 
\object{HD\,27442} is amongst the lowest
temperature stars in our sample. Its position in the Be vs. T$_{\mathrm{eff}}$
diagram follows the main trend. On the other hand, \object{HD\,38529} lies in
the Be gap for solar-temperature stars discussed in the previous section,
suggesting that the same mechanism that was able to deplete
Be in the dwarfs that are in this temperature interval may also
have played some role here. Curiously, these are the two sub-giants with the lowest absolute magnitude
in the HR diagram of Fig.\,\ref{fig_hr}. 
\citet[][]{Pri98} have identified a sub-giant (\object{HD\,160617}) in the same 
temperature regime as the objects discussed in Sect.\,\ref{sec:gap} that 
is Li normal, but Boron and possibly Be depleted. Given the low metallicity of this star 
([Fe/H]=$-$1.8), however, the authors have proposed that the observed trend has
to do with the low Boron and Be abundances existent in the Galactic region
where (and when) this star was formed. 
In general, none of the stars mentioned above seems to defy 
light-element depletion models, in the sense that their light-element depletion pattern
is consistent with Li being burned faster than Be.

{\it \object{HD\,10697} and \object{HD\,117176}}. 
These stars, both planet-hosts, seem to have preserved both their Li and Be abundances.
However, if on one hand they present ``normal'' Be abundance, their Li abundances are 
much higher than typically found for stars of their temperature. An extensive discussion of these two stars
was already done in \citet[][]{San02}. The most convincing explanation
found is that these two objects have just left
the main sequence, where they were hotter than they are now, 
but that there has not yet been time for their Li (and Be) to become strongly depleted.
Indeed, a look at Fig\,\ref{figbeteffgiants} shows that dwarfs that are merely
250\,K hotter are able to maintain high levels of Li and Be in
their atmospheres. Alternative explanations for the observed trend imply
a scenario of planetary material engulfment, or that the Li and Be ``excesses'' observed 
are due to a dredge up effect from a ``buffer" below the former main sequence convective 
envelope \citep[][]{Del90}. It should be mentioned, however, that
this latter scenario does not seem to be supported by current 
observations \citep[e.g.][]{Ran99}. For a more thorough discussion we point the 
reader to \citet[][]{San02}.

{\it \object{HD\,168443}}. This sub-giant is probably the least evolved
star amongst these 6, as attested by it surface gravity value of $\log{g}$=4.22,
and the relatively small height above the main sequence (see Fig.\,\ref{fig_hr}).
Its position in both the Li and Be vs. T$_{\mathrm{eff}}$ plots is that expected for
stars of its temperature. Curiously, its position in the HR diagram of Fig.\,\ref{fig_hr}
is very similar to those of \object{HD\,10697} and \object{HD\,117176}, although the
three stars present quite different Be abundances.

\begin{figure}[t]
\psfig{width=\hsize,file=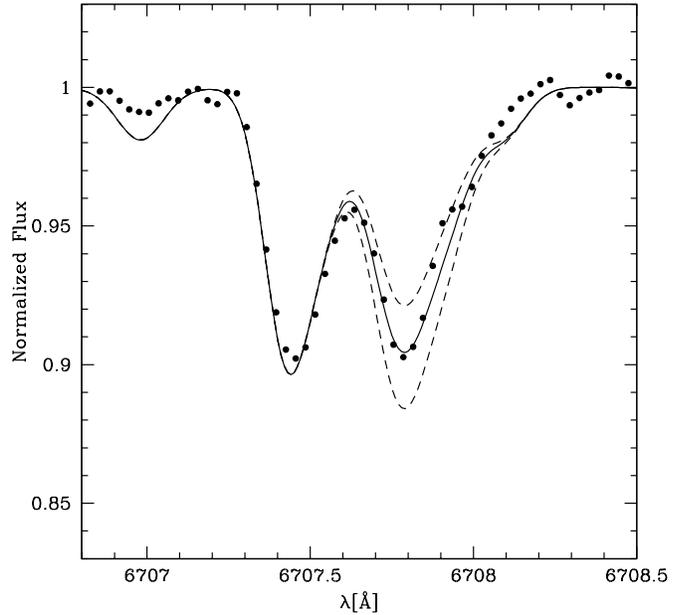}
\caption[]{Observed CORALIE spectrum of \object{HD\,23249} in the region near the Li6708\AA\ line 
(points) together with 3 spectral syntheses for abundances of $\log{N(Li)}$=1.14, 
1.24 (best fit) and 1.34\,dex. The spectral syntheses were done using the detailed line list
used in \citet[][]{Isr01}. The small inconsistencies in the fit are probably due to the low temperature of
this star, while the line list was optimized for the Sun. The strength of the Li line 
together with the quality of the line list used gives us a guarantee that the observed 
line is indeed due to Li and not any other element. }
\label{figli23249}
\end{figure}

{\it \object{HD\,23249}}. This ``single'' star (not known to have any
planetary companion), one of the coolest and most evolved ($\log{g}$=3.77) 
objects in our sample, is probably the most enigmatic case. Although its Be 
abundance is clearly low (only an upper limit was found -- see also Fig\,\ref{figgiant}), 
its Li content seems to be particularly high for its temperature, and a clear Li 
detection was possible for this star (see Fig.\,\ref{figli23249}). Considering the upper 
limit derived for the Be abundance $\log{N(Be)}$=0.15, and the value of $\log{N(Li)}$=1.14,
this star is depleted in Be by at least a factor of 13 (considering an initial Be
abundance of 1.26 -- see above), and its Li is depleted by a much larger factor of 150, considering
initially a meteoritic Li abundance of 3.31 \citep[][]{And89}.
For this case to be consistent with the expected light-element depletion pattern, 
the existence of Li in this cool
(old) sub-giant leads us to suppose that a process similar to the one discussed
above for \object{HD\,10697} and \object{HD\,117176} has occurred,
which is more difficult in this case, since this star is much cooler.
An alternative possibility to explain the observed abundances would be to consider that
the Li ``excess'' observed is due to a dredge up effect from a ``buffer" below the former
main sequence convective envelope \citep[][]{Del90,Ran99}, or to the accretion of metal-rich material
into the convective envelope of this star. At this moment the results are not conclusive, and
the star remains enigmatic.\\

\section{``Li-rich'' dwarfs and simultaneous Li and Be depletion}
\label{sec:lirich}

\begin{figure}[t]
\psfig{width=\hsize,file=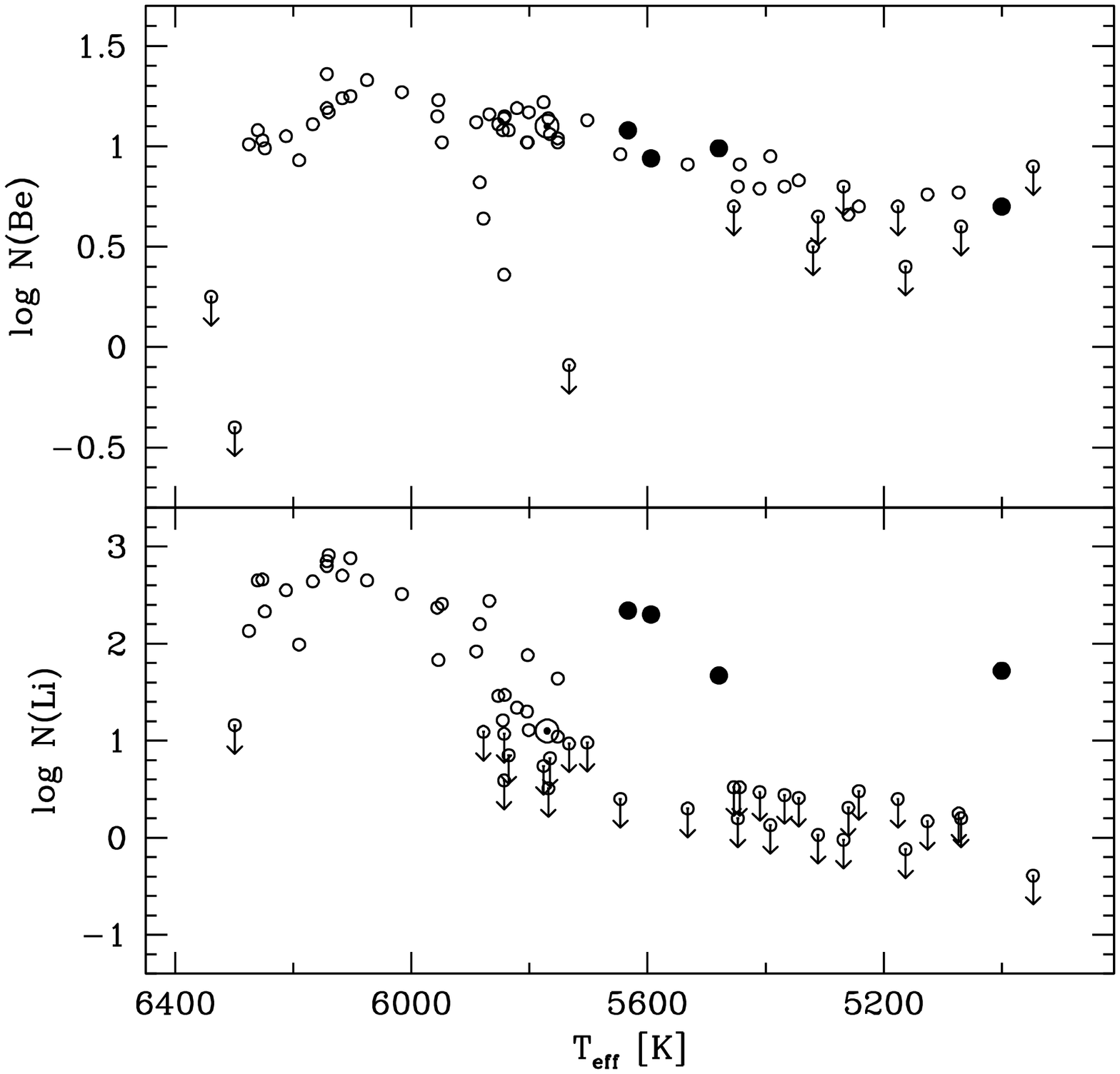}
\caption[]{Beryllium (upper panel) and Li (lower panel) abundances as a function 
of effective temperature for the dwarfs in our sample. The 4 dwarfs with effective temperatures below 
5700\,K that fall outside of the trend in the Li vs. T$_{\mathrm{eff}}$ plot are denoted by the 
filled circles. }
\label{figbeteffhighli}
\end{figure}

\begin{figure*}[t]
\psfig{width=\hsize,file=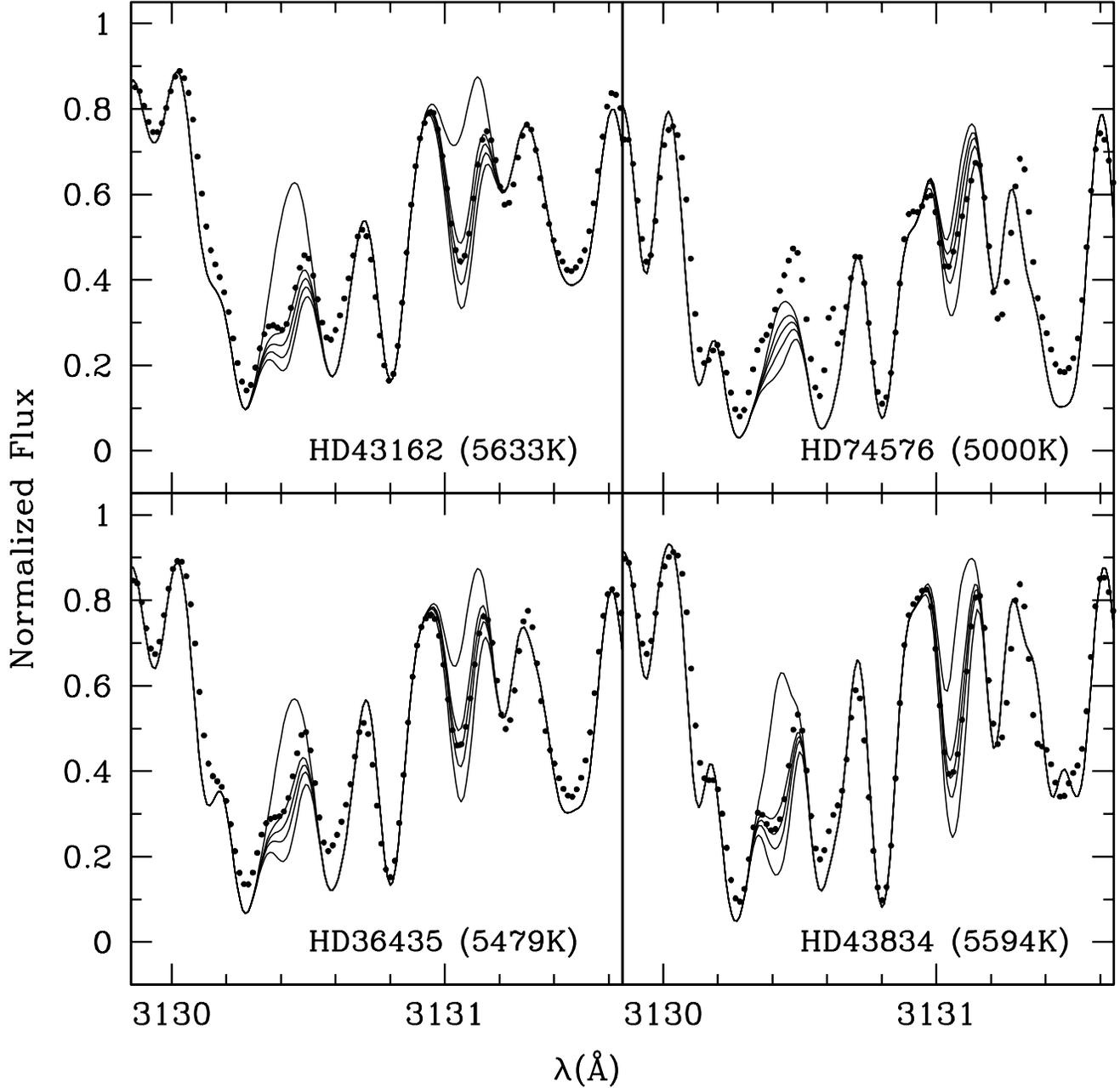}
\caption[]{Spectral syntheses (continuous lines) and observed spectra (points) in 
the \ion{Be}{ii} line region for the four stars discussed in Sect.\,\ref{sec:lirich}. In all panels, 
the upper and lower syntheses were done with a $\log{N(Be)}$ of 1.42 (meteoritic) and $-$10.0 
(essentially no Be), respectively. Except for \object{HD\,74576}, the three other spectral 
syntheses correspond to the optimal fit and to fits with abundance variations 
of $\pm$0.15\,dex. For \object{HD\,74576}, these three intermediate fits correspond to the 
optimal fit and to fits with abundance variations of $\pm$0.30\,dex.}
\label{fighighli}
\end{figure*}

In Fig.\,\ref{figbeteffhighli} there are 4 dwarfs (\object{HD\,36435}, \object{HD\,43162}, 
\object{HD\,43834}, and \object{HD\,74576}) with effective temperatures below 
5700\,K, that have measurable Li abundances, and clearly fall off the main trend (see also Fig.\,\ref{fighighli}). 

To investigate their status we have tried to derive their ages from measurements 
of the chromospheric activity S index. Values of S were found in the literature for \object{HD\,36435} 
(S=0.40), \object{HD\,43834} (S=0.18), and \object{HD\,74576} (S=0.74) \citep[][]{Hen96}. 
For \object{HD\,43162} we have derived S=0.41, on the Mount-Wilson scale, based on 
CORALIE spectra using the procedure described in \citet[][]{San00}. 
These values imply ages below 1\,Gyr for all the stars \citep[][]{Hen96}, with the exception of
\object{HD\,43834}. 

The S index derived by \citet[][]{Hen96} for this latter dwarf is
based on one single measurement, while from 8 different CORALIE spectra we have
derived a higher value of S=0.33, a value implying an age
below $<$1\,Gyr (for \object{HD\,36435} and \object{HD\,74576} our CORALIE spectra
give S values of 0.49 and 0.69, respectively, compatible with the results
derived by \citet[][]{Hen96}). However, doubt is also cast on the young 
status of \object{HD\,43834} as a young star by its X-ray luminosity 
\citep[L$_{\mathrm{x}}$=2.8\,10$^{27}$\,erg\,s$^{-1}$ --][]{Hue98}, a value not
characteristic of a young object.

A look at Table\,\ref{table1} also reveals that the other three stars have 
projected rotational velocities above 3.5\,km\,s$^{-1}$.
In other words, except perhaps for \object{HD\,43834}, these objects seem to be quite young, 
which probably justifies their higher than average Li contents.

\object{HD\,43834} thus probably remains the only problematic case. If it is an ``old'' star,
how can we explain that at $\sim$5600\,K it still has some Li in its atmosphere, clearly
failing to follow the trend of the remaining dwarfs of its temperature? A possible explanation
would be that this star has recently engulfed planetary (Li-rich, and H-poor)
material. However, at this moment we do not have any possibility of confirming this. 
In any case, this star is not known to be orbited by any planetary-mass
companion (see Paper\,A). Another possible (but probably less likely) explanation would be 
that the interstellar material from which this star was formed was particularly
Li-rich.

Curiously, inspection of Fig.\,\ref{figbeteffhighli} and Table\,\ref{table1} tells us
that all these stars already seem to have depleted some of their Be, while there is a slight impression 
that they may lie near the upper envelope of the $\log{N(Be)}$ vs. T$_{\mathrm{eff}}$ trend
(i.e. these stars may have burned less Be than their older counterparts).
This seems to tell us that Be depletion in late-type dwarfs stars starts at early ages (i.e. before 1\,Gyr).

These four stars are thus good examples of cool dwarfs that have burned Be but still 
have large detectable quantities of Li in their atmospheres.
A curious case among these objects is \object{HD\,74576}, the coolest and probably 
youngest star (based on its chromospheric activity) of the four. If its Li and Be contents are confirmed,
Li is depleted by a factor of $\sim$40, while Be is depleted by a factor of $\sim$4, 
considering initial values for Li and Be of 3.31 and 1.26\,dex, respectively. The 
process responsible for this simultaneous Li and Be depletion might be similar to the 
one found for F dwarfs \citep[e.g.][]{Del97} -- see Sect.\,\ref{sec:libe}.

\section{Lithium vs. Beryllium}
\label{sec:libe}

\begin{figure}[t]
\psfig{width=\hsize,file=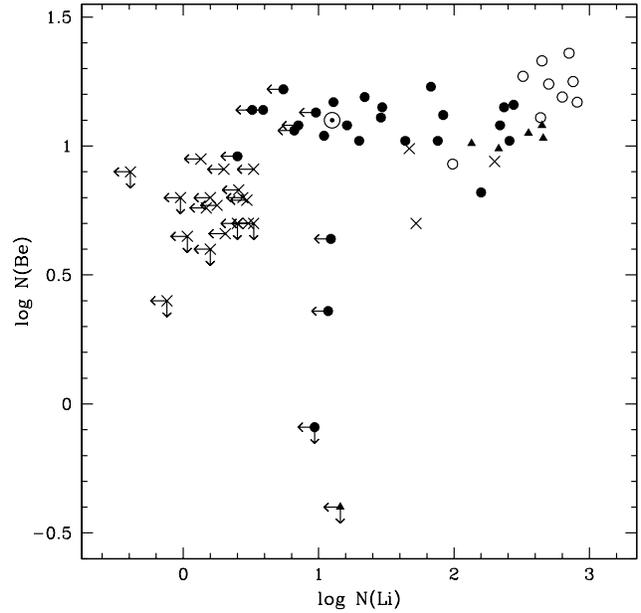}
\caption[]{Li and Be abundances plotted against each other. Only dwarfs are considered in this
plot. Triangles represent stars with T$_{\mathrm{eff}}$$>$6200\,K, open circles stars with 6000$<$T$_{\mathrm{eff}}$$\leq$6200\,K, dots objects with
5600$<$T$_{\mathrm{eff}}$$\leq$6000\,K, and crosses stars with T$_{\mathrm{eff}}$$\leq$5600\,K. }
\label{figlibe}
\end{figure}

In Fig.\,\ref{figlibe} we plot the Li and Be abundances against each other. 
For simplicity, only dwarfs were considered in this plot. The figure tells
us that, overall, there seems to be a correlation between the depletions
of Li and Be, in the sense that stars that have depleted their Be 
have also strongly depleted their Li. This trend was indeed already clear from 
e.g. Fig.\,\ref{figbeteffhighli}.

In Fig.\,\ref{figlibe} dwarfs in different temperature regimes are denoted
by different symbols. Triangles represent stars with T$_{\mathrm{eff}}$$>$6200\,K, 
open circles stars with 6000$<$T$_{\mathrm{eff}}$$\leq$6200\,K, dots objects with
5600$<$T$_{\mathrm{eff}}$$\leq$6000\,K, and crosses denote stars with T$_{\mathrm{eff}}$$\leq$5600\,K. 

Excluding the first group of objects, which lies 
near the Li and Be gap for F stars and is clearly depleted in both Li and Be, we can 
see a clear progression in the Li/Be ratios as a function of temperature. For stars 
with 6000$<$T$_{\mathrm{eff}}$$\leq$6200\,K, depletion of Be
has not occurred, while Li is moderately depleted (the particularly high
Be abundances for these stars might be also due to Galactic evolution effects, since these objects 
are on average quite metal-rich -- see Sect.\,\ref{sec:beteff}). 
Interestingly, there is some hint of a correlation between Li and Be abundances
for objects with temperatures above $\sim$6000\,K -- see also e.g. Fig.\,\ref{figbeteffhighli}.
In the temperature regime between 5600 and 6000\,K, we see a large dispersion in the Li
abundances, while the Be abundances remain very close to the primordial value. 
On the other hand, for temperatures below 5600\,K, both Li and Be are clearly depleted. 
This temperature seems to mark the onset of strong Be depletion in solar-type stars.
Overall, and as already discussed in \citet[][]{San02}, the plot of Fig.\,\ref{figlibe}
gives us information about the temperatures for which Li and Be depletion starts. 

Although almost all the stars agree with this picture, a few deserve particular 
comment. The three dwarfs discussed in Sect.\,\ref{sec:gap}, all in the 
temperature range between 5600 and 5900\,K -- filled circles -- (\object{HD\,4391}, 
\object{HD\,20766}, \object{HD\,20807}, and maybe also \object{HD\,114762}), all fall 
clearly off the general trend in Fig.\,\ref{figlibe}. This result seems to indicate
that some kind of non-standard Be and Li burning mechanism has occurred for
these dwarfs and is responsible for the observed Be-gap for solar
temperature stars. 


Three other objects, in the temperature regime below 5600\,K (crosses), also seem to
be particularly Li-rich. These stars, \object{HD\,36435}, \object{HD\,43834}, 
\object{HD\,74576}, were already discussed in Sect.\,\ref{sec:lirich}. Except maybe for
\object{HD\,43834}, they are probably young stars, and as mentioned in Sect.\,\ref{sec:lirich}
their abnormally high Li abundances are probably explained by their young ages. However, Fig.\,\ref{figlibe}
also implies that Be depletion has already occurred for these dwarfs, and
is particularly severe for the case of \object{HD\,74576} (the coolest of
the stars discussed in Sect.\,\ref{sec:lirich}). This result strongly supports the
conclusion that for very cool solar-type dwarfs, Be depletion occurs at 
high rates early during their lifetimes. 

Interestingly, the position of \object{HD\,36435}, \object{HD\,43834} and \object{HD\,74576} 
in the Li vs. Be diagram (the three crosses with the highest Li abundances in Fig.\,\ref{figlibe})
are reasonably well aligned with the points representing stars with T$_{\mathrm{eff}}$$>$6000\,K 
(open circles and triangles in the same figure). Although preliminary and based on three stars only, this might suggest that a similar mechanism may have been responsible for the depletion of Li and Be in both groups of stars.

\section{Concluding remarks}

In this paper we have analyzed the Be abundances for a large sample of field solar-type stars 
with temperatures in the range $\sim$4800-6300\,K. This sample gives us the possibility
to study the depletion of Be for dwarfs and sub-giants of different temperature.
The major results obtained can be summarized as follows.

\begin{itemize}

\item A plot of Be as a function of T$_{\mathrm{eff}}$ shows that the abundances
of this light element have a maximum near 6100\,K, decreasing both towards
higher temperatures (towards the already known Be-gap for F
stars), and lower temperatures. This latter trend \citep[see also][]{San02},
similar to the one found for Li (that also peak near 6100\,K), is particularly evident for
temperatures below 5600\,K. We have explored several possible biases that could
lead to the observed behaviour, and found that none could satisfactorily explain it.
We thus conclude that this result probably reflects a disagreement between the
observations and the models dealing with Be depletion including rotational mixing 
mechanisms \citep[e.g.][]{Pin90,Del93}. A solution may involve the inclusion of 
internal wave-driven mixing mechanisms in the models \citep[e.g.][]{Mon00}. The 
fact that literature data for relatively young clusters do not seem do present 
the same clear trend might indicate that stars cooler than this value do deplete 
Be during their main-sequence evolution. 

\item We have found 4 dwarfs (\object{HD\,4391}, \object{HD\,20766}, 
\object{HD\,20807}, and \object{HD\,114762}) and one sub-giant star (\object{HD\,38529}), with 
temperatures between 5600 and 5900\,K, that present particularly low Be abundances, indicating
the presence of some kind of a Be-gap for solar-temperature stars. While for \object{HD\,114762} this result
can be explained as due to Be Galactic evolution effects (this star is reasonably
metal-poor), for the other objects the low Be-abundances imply that some unknown extra-mixing
process has occurred. This result is not predicted by any current Be depletion model.

\item A cool sub-giant (\object{HD\,23249}) star was discovered for which only an upper limit for 
the Be abundance was measured, but that presents clearly detectable amount of Li in 
its atmosphere. We have no clear explanation for this, although
some possibilities, like the accretion of planets, or the dredge-up of Be
from a ``buffer'', may account for the observed Li and Be contents.
Other sub-giants in our sample present ``normal" Be abundances (they
follow the general trend observed for dwarfs), although two of them (HD\,10697 and HD\,117176) 
also have particularly high Li values.

\item We have found four dwarfs with temperatures below 5700\,K with particularly high Li abundances,
all presenting signs of Be depletion. At least three of these (\object{HD\,36435}, \object{HD\,43162}, 
and \object{HD\,74576}) are young, while another (\object{HD\,43834}) is probably older 
than 1\,Gyr, making its Li and Be content quite difficult to explain. For the three younger 
dwarfs, the ``depleted'' Be pattern suggests that the depletion of this light element 
occurs at early ages in the coolest solar-type dwarfs (this is particularly evident for
\object{HD\,74576}). A comparison of Li and Be abundances also suggests that the 
mechanism responsible for 
this depletion may be similar to the one found for the hotter (T$_{\mathrm{eff}}$ 
above $\sim$6100\,K) F-dwarfs.

\item A comparison of Li and Be suggests that their abundances follow a clear
progression. For stars with temperatures above $\sim$6200\,K, these light elements are
both depleted, in the well known Li and Be-gaps for F stars. A maximum in the abundances
of Li and Be is found for temperatures between 6000 and 6200\,K, although for
Be this maximum may be due to stellar population bias in our sample. For cooler 
dwarfs Li is severely depleted, while Be may present a plateau down to a temperature 
of $\sim$5600\,K that marks the onset of severe Be depletion.

\end{itemize}

These results seem to defy the current models of Be depletion in solar-type stars.
In this respect, of particular interest is the discovery of a class of severely 
Be-depleted solar-type stars, that form some kind of Be-gap, and the suggestion that Be burning
does occur for dwarfs with temperatures below $\sim$5600\,K. 
Although preliminary, our data also suggests that Be depletion may occur before or on the
main sequence for stars in this temperature regime.
To better understand the observed discrepancies it would be important
to enlarge our samples, and to obtain precise Boron abundances for at least
the most problematic cases.

\begin{acknowledgements}
  We would like to thank the anonymous referee for the careful
  and constructive report. Support from Funda\c{c}\~ao 
  para a Ci\^encia e Tecnologia (Portugal) to N.C.S. in the form of 
  a scholarship is gratefully acknowledged.
\end{acknowledgements}

\end{document}